\documentclass[twocolumn,twoside,slac]{revtex4}
\usepackage{graphicx}
\usepackage{fancyhdr}
\renewcommand{\headrulewidth}{0pt}
\renewcommand{\footrulewidth}{0pt}

\begin{document}

\title{IceCube's Development Environment.}

%

\author{S. Patton}
\affiliation{LBNL, Berkeley, CA 94720, USA}
\author{D. Glowacki}
\affiliation{Space Science and Engineering Center, University of
Wisconsin, Madison, WI 53706, USA}

\begin{abstract}
When the IceCube experiment started serious software development it
needed a development environment in which both its developers and
clients could work and that would encourage and support a good software
development process. Some of the key features that IceCube wanted in
such a environment were: the separation of the configuration and build
tools; inclusion of an issue tracking system; support for the Unified
Change Model; support for unit testing; and support for continuous
building. No single, affordable, off the shelf, environment offered all
these features. However there are many open source tools that address
subsets of these feature, therefore IceCube set about selecting those
tools which it could use in developing its own environment and adding
its own tools where no suitable tools were found. This paper outlines
the tools that where chosen, what are their responsibilities in the
development environment and how they fit together. The complete
environment will be demonstrated with a walk through of single cycle
of the development process.
\end{abstract}

\maketitle

\thispagestyle{fancy}


\section{Introduction}

\subsection{IceCube Overview}

The IceCube experiment is a neutrino telescope in which 1~km$^{3}$ of
ice is instrumented with Digital Optical Modules (DOMs), i.e. light
detectors. The experiment is comprised of around 150 collaborators from
all over Europe and America, but it is located at the South Pole. This
location provides some fairly unique problems such as the following:

\begin{itemize}
\item For 8 months of the year the Pole is manned by ``Winter-overs''
only and there is no outside access.
\item These ``Winter-overs'' are not usually the same people who develop
the experiment's software.
\item The satellite connection to the Pole is limited and can also be
intermittent.
\end{itemize}

Given these conditions not only is software reliability a {\em major}
asset to the experiment, but it is also essential for the
``Winter-overs'' to be able to handle any software issue that arise
during the 8 months that the Pole is shut. Therefore it is important for
IceCube to have a software development environment that can help
mitigate these issues.

\section{Software Development Environment}

\subsection{Requirements}

The software reliability issue led IceCube to make the decision that
one of its key requirements for its software development environment is
that it must support and encourage industry proven ``good practices''.
This was considered to be one of the best ways to mitigate the risk to
the whole project contributed by the software reliability issue. This
led to the environment needing to provide tools which covered the
following practices:

\begin{description}
\item[Work Space Management] This includes support for some version of
the ``Unified Change Management'' model~\cite{ucm-whitepaper}.
\item[Code Building] This is separate for the work space management, as
this tools will have language dependent portions, whereas work space
management should be language neutral.
\item[Unit Testing] This helps capture the requirements of the software
in a concrete form. It also means that software units can be replaced
provided that the new code passes these tests.
\item[Continuous integration] This covers the repeated building, testing
and reporting of any new code additions, so that errors can be found
early in the development process.
\item[Issue Tracking] This provides an institutional memory for the
experiment as there is expected to be significant turnover of personnel
due to the long lifetime of the experiment.
\end{description}

Another decision, that also addressed reliability risk, was that as much
software as possible at the Pole should be written in Java as this was a
found to be a more robust language than C++. However the physicists
required that their online filter code be written in C++ so that they
would be able to exploit the resources provided by the ROOT
package~\cite{root-site}. Similarly, DOM programming is best done in C,
due to the nature of the hardware. This meant the the development
environment needed multi-language support.

\begin{figure*}
\centering
\includegraphics{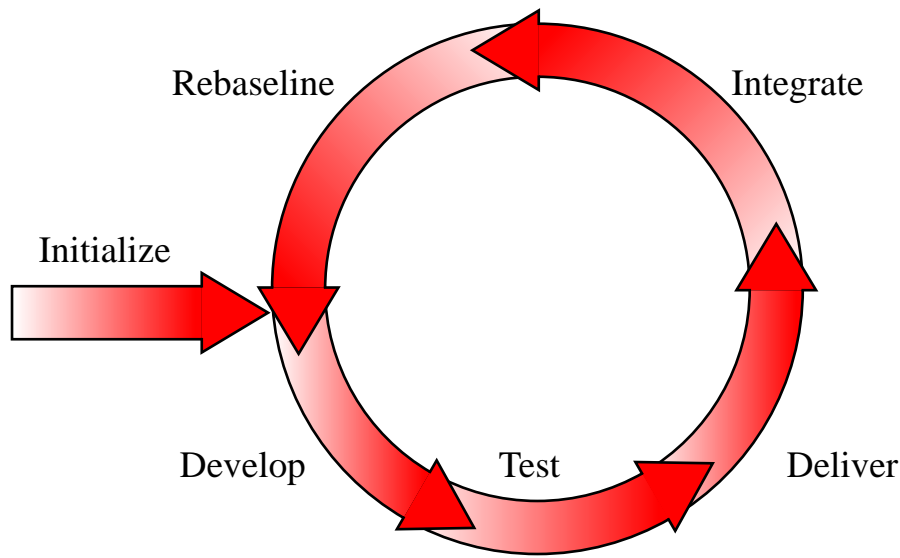}
\caption{The work space management cycle as defined by the Unified
Change Management model.} \label{ucm-cycle}
\end{figure*}

\subsection{Implementation}

No single, affordable, off the shelf product was found that fulfilled
all these requirements. Meanwhile it was observed that there were many
open source tools that addressed subsets of these requirements.
Therefore IceCube decided to create its own development environment by
selecting those open source tools which it could exploit and develop its
own tools in those areas where no open source solution existed. In the
end this turned out to be only one area, work space management.

\subsubsection{Work Space Management}

Figure~\ref{ucm-cycle} illustrates the basic ideas behind the work space
management cycle as defined by the Unified Change Management model. A
developer enters the cycle by initializing an area to be a work space to
tack a specified issue. Once the space is established, the developer is
free to work on the code and its test, including adding new code and new
tests if the issue is an enhancement request. Eventually the point
should be reached when the change is notional complete. At this point in
time the code should pass all of its unit tests. Once this
requirement has been satisfied the code is then delivered, ready for
integration with the rest of the software product.

The exact results of code delivery depends a great deal on the size of
the product. For small products delivery may simply be inclusion of the
new and modified code in the next continuous integration test of the
product. Large projects, on the other hand, will be broken down into
hierarchical subsystems and ``delivery"" will mean that the new and
modified code is passed on to an integrator who will bring together all
the changes in that subsystem, test that they work together and then
pass the resulting code base on to the next level of the hierarchy.

As noted above, no open source tool was found to handle these
responsibilities so the {\tt bfd} tool has been created. This
initializes work spaces with the files and soft links they need to be
able to work. It also adds a layer of policy on top of a code archive
system, in IceCube's case this is CVS, that handles code checkout,
checkin and tagging.

\begin{figure*}[t]
\centering
\begin{verbatim}
Main targets:

clean            Clean all directories and files built for this project
compile          Compile this project
createClass      Create new .java files for a class an its matching test
createInterface  Create new .java files for an interface an its matching
                 test
createPackage    Create a new package
docs             Create API and test documents for this project
javadocs         Create the Javadocs for this project
lib              Create the library for this project
report           Create the report on the tests run for this project
test             Run the tests for this project
\end{verbatim}
\caption{The public targets for the standard {\tt build.xml} file for a
project.}
\label{std-build-xml}
\end{figure*}

\subsubsection{Code Building}

As the core of the IceCube software system will written be in Java, it was
decided that {\tt ant}~\cite{ant-site} would be the best choice as the
main build tool for the environment. This tool is already the standard
build tool for many Java products. One of the major benefits is that it
is written in Java which means that it can execute anywhere you expect
Java code to execute.

Also, by being written in Java, {\tt ant} easily extensible although
there are already plenty of ``tasks'' for handling the more common duties
associated with code building such as; skeleton processing; running {\tt
JUnit} (see below); integration with other tools such as {\tt
JDepend}~\cite{jdepend-site} - a metric calculation program.

The execution of {\tt ant} is driven by a {\tt build.xml} file,
therefore a standard {\tt build.xml} file is provided for each project
within the IceCube software system. Figure~\ref{std-build-xml} shows the
public targets for the standard file.

Similarly a standard {\tt build.xml} file is provided for each work
space. This file contains call downs to the public targets of each
project contained in a work space, as well as extra targets that handle
work space level issues. e.g. creation of deployment tar-balls.

\subsubsection{Unit Testing}

The model adopted for unit testing in IceCube is the XUnit
framework~\cite{xunit-paper}. The Java implementation of this framework
is called JUnit~\cite{junit-site}. In this framework a set of test for a
particular unit (at the base level this is a class) are created and
grouped together in a ``TestCase'' for that unit. Each test codifies one
of the requirements that has been specified for that unit.

A skeleton for the TestCase for a class is generated by {\tt ant} when
{\tt ant} is used to generate the class skeleton. Thus, for the simpler
classes the developer simply needs to provide a set of implementations
that test all requirements and the job is complete.

{\tt JUnit} provides both a GUI and text based interface for testing.
However {\tt ant} also provides its own interface and this is the one
used by IceCube. The {\tt ant} interface generates an XML file as well
as providing a summary of the execution of the tests. A further {\tt
ant} task can be used to convert this XML file into a set of HTML frames
that can easily be viewed to see which tests succeeded and which failed
and why.

\subsubsection{Continuous integration}

The earlier a problem can be discovered the sooner it can be fixed and
the less impact it will have on the overall development of a software
product. To that end it makes sense to continuously integrate all parts
of a product to check that no change has broken another part of the
software. IceCube's solution to this issue is to use the
CruiseControl~\cite{cc-site} software. This can be integrated with {\tt
ant} build files so is a good match to the rest of the IceCube system.
Builds are currently executed every time a change committed to the CVS
archive and a build is not in progress. Work is in progress to also
allow builds to be scheduled, e.g. every night.

The results of a build can be readily made available on the Web, along
with the results of the unit test.

As well as the standard CruiseControl output, the results can be passed
to the Tinderbox~\cite{tinder-site} program which can display a graphical
of the current state of various different configurations of builds.

\subsubsection{Issue Tracking}

In the initial incarnation of the IceCube development environment
Issuezilla was chosen to handle issue tracking. This product was an
outgrowth of the Bugzilla product used by Netscape to track Mozilla
issues. However the future of this product is unclear as its developers
are now focused on a Java based replacement, Scarab, whose development
appears to a stalled over the last half year of so.

In the meantime an open source development of the old SourceForge
project, GForge~\cite{gforge-site}, has recently announced its presence.
This tool not only provides issue tracking facilities, but it can also
act as a ``portal'' though which the entire software effort of IceCube
could be accessed.

All this means that the Issue tracking portion of the IceCube
development environment has yet to be finally settled.

\section{Walk Through of a User Session}

In this section of the paper we will see an example walk through of the
development environment highlighting many of the typical uses. Some
features of the environment, e.g. baselining, are not yet implemented
and so are not included in this example.

\begin{figure*}[t]
\centering
\begin{verbatim}
[patton@glacier patton]$ mkdir work
[patton@glacier patton]$ cd work
[patton@glacier work]$ bfd init /home/icecube/tools
  ... <output skipped> ...
[patton@glacier work]$ ls -l
total 24
-rw-rw-r--    1 patton   patton        297 Jun  9 15:31 build.xml
-rw-rw-r--    1 patton   patton       5443 Jun  9 15:31 setup.csh
-rw-rw-r--    1 patton   patton       5065 Jun  9 15:31 setup.sh
drwxrwxr-x    6 patton   patton       4096 Jun  9 15:31 tools
[patton@glacier work]$ source setup.sh
\end{verbatim}
\caption{An example of work space creation.}
\label{ws-create}
\end{figure*}

\subsection {Creating a Work Space}

Figure~\ref{ws-create} shows an example of how a work space can be
initialized. It is fairly straightforward. A directory, which will be
the work space, is creates and then {\tt bfd~init} is executed in that
directory. This created the necessary files and soft links. The example
shows the contents of the work space after it is initialized.

The final set of the example shows the ``sourcing'' of the appropriate
{\tt setup} file that sets up the necessary environmental variables to
use the work space. This command needs to be executed every time a
session begins work in the work space.

\subsection{Checking out and Building a Project}

Figure~\ref{co-and-build} shows how to checkout a project, in this case
the {\tt icebucket} project, from the code archive and then build it
using the {\tt ant} command. As you can see in the example two jar files
are created, one holds the project code, and the other holds the code
and resources needed to run the unit tests of the project.

\begin{figure*}[t]
\centering
\begin{verbatim}
[patton@glacier work]$ bfd co icebucket
  ... <output skipped> ...
[patton@glacier work]$ cd icebucket 
[patton@glacier icebucket]$ ant
Buildfile: build.xml
  ... <output skipped> ...
BUILD SUCCESSFUL
Total time: 9 seconds
[patton@glacier icebucket]$ cd ..
[patton@glacier work]$ ls -l lib 
total 8
-rw-rw-r--    1 patton   patton       3455 Jun  9 15:39 icebucket.jar
-rw-rw-r--    1 patton   patton       3004 Jun  9 15:39 icebucket-test.jar
\end{verbatim}
\caption{An example of checking out and building the {\tt icebucket}
project.}
\label{co-and-build}
\end{figure*}

\subsection{Creating a New Project}

Figure~\ref{new-project} shows how the contents of a new project, {\tt
gromit}, can be created using {\tt ant}. Dependencies of one project on
another are stored in the {\tt project.xml} file of project, thus the
need to edit this file in the example to include the fact that {\tt
gromit} is dependent upon {\tt icebucket}.

\begin{figure*}[t]
\centering
\begin{verbatim}
[patton@glacier work]$ bfd co gromit
  ... <output skipped> ...
[patton@glacier work]$ ant -DPACKAGE=icecube.tools.examples \
>         -DPROJECT=gromit createProject
Buildfile: build.xml
  ... <output skipped> ...
BUILD SUCCESSFUL
Total time: 2 seconds
[patton@glacier work]$ ls -l gromit 
total 16
-rw-rw-r--    1 patton   patton        397 Jun  9 15:45 build.xml
-rw-rw-r--    1 patton   patton        292 Jun  9 15:45 project.xml
drwxrwxr-x    4 patton   patton       4096 Jun  9 15:45 resources
drwxrwxr-x    3 patton   patton       4096 Jun  9 15:45 src
[patton@glacier work]$ emacs gromit/project.xml
  ... <add dependency on icebucket> ...
[patton@glacier work]$ bfd uadd gromit
\end{verbatim}
\caption{An example of populating a new project with its default files.}
\label{new-project}
\end{figure*}

\subsection{Creating a New Class}

Figure~\ref{new-class} shows how {\tt ant} can be used to create new
class files. It also shows that the test file for a class is also
created at the same time as the class itself.

\begin{figure*}[t]
\centering
\begin{verbatim}
[patton@glacier work]$ cd gromit
[patton@glacier gromit]$ ant -DCLASS=Counter2 createClass
  ... <output skipped> ...
[patton@glacier gromit]$ ls -lR src/icecube/tools/examples/ 
src/icecube/tools/examples/:
total 12
-rw-rw-r--    1 patton   patton       1008 Jun  9 15:50 Counter2.java
-rw-rw-r--    1 patton   patton        575 Jun  9 15:45 package.html
drwxrwxr-x    2 patton   patton       4096 Jun  9 15:50 test

src/icecube/tools/examples/test:
total 8
-rw-rw-r--    1 patton   patton       2178 Jun  9 15:50 Counter2Test.java
-rw-rw-r--    1 patton   patton        424 Jun  9 15:45 package.html
\end{verbatim}
\caption{An example of creating a skeleton file for a new class and its
tests}
\label{new-class}
\end{figure*}

\subsection{Implementing a Class}

Figure~\ref{impl-class} shows the steps that can now be taken to
provide an implementation of a class and its unit test. It also shows
how a {\tt JUnit} test can be run outside {\tt ant}.

\begin{figure*}[t]
\centering
\begin{verbatim}
[patton@glacier gromit]$ cd src/icecube/tools/examples/
[patton@glacier examples]$ emacs test/Counter2Test.java
  ... <write tests> ...
[patton@glacier examples]$ emacs Counter2.java 
  ... <implement class> ...
[patton@glacier examples]$ cd ../../../..
[patton@glacier gromit]$ ant lib
  ... <output skipped> ...
[patton@glacier gromit]$ cd ..
[patton@glacier work]$ java -cp lib/gromit-test.jar \
> junit.textui.TestRunner icecube.tools.examples.test.Counter2Test
.....
Time: 0.159

OK (5 tests)
\end{verbatim}
\caption{An example showing the steps used to implement a new class and
its tests.}
\label{impl-class}
\end{figure*}

\subsection{Delivering a Project}

Figure~\ref{delivery} shows how, once development of a class has been
completed, it can be delivered and thus made available to the next stage
of integration.

\begin{figure*}[t]
\centering
\begin{verbatim}
[patton@glacier work]$ bfd uadd gromit
  ... <output skipped> ...
[patton@glacier work]$ bfd archive -m "New example project" gromit 
  ... <output skipped> ...
[patton@glacier work]$ bfd deliver -j gromit
Are you sure you want to deliver "gromit" with tag V01-00-00
y/n: y
V01-00-00 of "gromit" has been delivered.
\end{verbatim}
\caption{An example delivering a project after its modifications have
been completed.}
\label{delivery}
\end{figure*}

\subsection{Cleaning up}

Figure~\ref{cleaning} shows how you can clean up a work space once its
task has been accomplished. Using this command, rather than simply using a
{\tt rm~-rf}, checks that everything has been safely stored in the code
archive.

\begin{figure*}[t]
\centering
\begin{verbatim}
[patton@glacier work]$ bfd dispose gromit
No files have been added to, or modified in, "gromit".
There are no unknown files in, "gromit".
Disposed of "gromit"
[patton@glacier work]$ bfd dispose
Are you sure you want to dispose of the entire workspace?
y/n: y
No files have been added to, or modified in, "icebucket".
There are no unknown files in, "icebucket".
Disposed of "icebucket"
Disposed of workspace files...anything left is your own problem.
[patton@glacier work]$ ls -l 
total 0
\end{verbatim}
\caption{An example of cleaning up a work space once its task has been accomplished.}
\label{cleaning}
\end{figure*}

\section{Summary}

IceCube has created a development environment for its software based,
where possible, on open source tools. The aim of this environment has
been to simplify the developers work load by integrating ``good
practices'' into the environment.  So far these goals have been achieved
and early adopters of the environment have been pleased by the ease it
has brought to their development process.


\begin{thebibliography}{9}   

\bibitem{ucm-whitepaper}
http://www.rational.com/products/whitepapers/415.jsp

\bibitem{root-site}
http://root.cern.ch/

\bibitem{ant-site}
http://ant.apache.org/

\bibitem{jdepend-site}
http://www.clarkware.com/software/JDepend.html

\bibitem{xunit-paper}
http://www.xprogramming.com/testfram.htm

\bibitem{junit-site}
http://junit.org

\bibitem{cc-site}
http://cruisecontrol.sf.net/

\bibitem{tinder-site}
http://www.mozilla.org/tinderbox.html

\bibitem{gforge-site}
http://www.gforge.org/

\end{thebibliography}
\end{document}